\documentclass[sigconf, 10pt,nonacm=true, authorversion]{acmart}

\usepackage{graphicx} 
\usepackage{amsmath}
\usepackage{xspace}
\usepackage{multirow}
\usepackage{enumitem}
\usepackage{caption}
\usepackage{subcaption}
\usepackage{mathtools}
\usepackage{makecell}
\usepackage{pifont}
\usepackage{url}
\usepackage{enumitem}
\setitemize{leftmargin=0.15in}

\allowdisplaybreaks
\newcommand{\hz}[1]{\textcolor{red}{#1}}

\newcommand{\pname}{ChatTracer\xspace}

\title{\pname: Large Language Model Powered Real-time Bluetooth Device Tracking System}
\author{Qijun Wang, Shichen Zhang, Kunzhe Song, and Huacheng Zeng\\
Department of Computer Science and Engineering, Michigan State University\\
\{qjwang, sczhang, songkunz, hzeng\}@msu.edu\\\bigskip}
\bigskip

\date{September 2023}

\begin{document}

\pagestyle{empty}

\begin{abstract}
Large language models (LLMs), exemplified by OpenAI ChatGPT and Google Bard, have transformed the way we interact with cyber technologies. 
In this paper, we study the possibility of connecting LLM with wireless sensor networks (WSN).
A successful design will not only extend LLM's knowledge landscape to the physical world but also revolutionize human interaction with WSN. 
To the end, we present \pname, an LLM-powered real-time Bluetooth device tracking system. 
\pname comprises three key components: 
an array of Bluetooth sniffing nodes, a database, and a fine-tuned LLM.
\pname was designed based on our experimental observation that commercial Apple/Android devices always broadcast hundreds of BLE  packets per minute even in their idle status. 
Its novelties lie in two aspects:
i) a reliable and efficient BLE packet grouping algorithm;
and
ii) an LLM fine-tuning strategy that combines both supervised fine-tuning (SFT) and reinforcement learning with human feedback (RLHF).
%
%
%
We have built a prototype of \pname with four sniffing nodes.
Experimental results show that \pname not only outperforms existing localization approaches, but also provides an intelligent interface for user interaction.



\end{abstract}

\maketitle

\vspace{-0.1in}
\section{Introduction}

The emergence of large language models (LLM) has ushered in a transformative era, revolutionizing the way we interact with technology and harnessing the power of natural language processing. 
%
%
To the best of our knowledge, LLM has not yet been combined with wireless sensor networks (WSN) 
\cite{naveed2023comprehensive, zhao2023survey, hadi2023survey}. 
Connecting these two worlds is appealing for two reasons. 
First, from the LLM's perspective, 
connecting with WSN will extend LLM's capabilities from generating knowledge-based information to providing \textit{fresh}, \textit{real-time} sensory information of our physical world.
Second, from the WSN's perspective, the use of LLM will transform the interaction between humans and WSN, making the sensory information more accessible and easier to comprehend for end users. 


In this paper, we present the first-of-its-kind study on connecting LLM with WSN, to understand the potential of LLM in processing sensory data from WSN and facilitating human interaction with WSN. 
Specifically, we introduce \pname, an LLM-powered real-time Bluetooth device tracking system.
\pname has an array of radio sniffing nodes deployed in the area of interest, which keep listening to the radio signals emitted by the Bluetooth devices in the proximity. 
\pname processes its received Bluetooth packets to extract their physical and payload features using domain knowledge. 
The extracted per-packet features are stored in a database and fed into an LLM to generate the human-like textual response to user queries. 

\vspace{-0.1in}
\section{\pname: System Architecture\!\!\!\!\!}
\begin{figure}
\centering
 \includegraphics[width=0.985\linewidth]{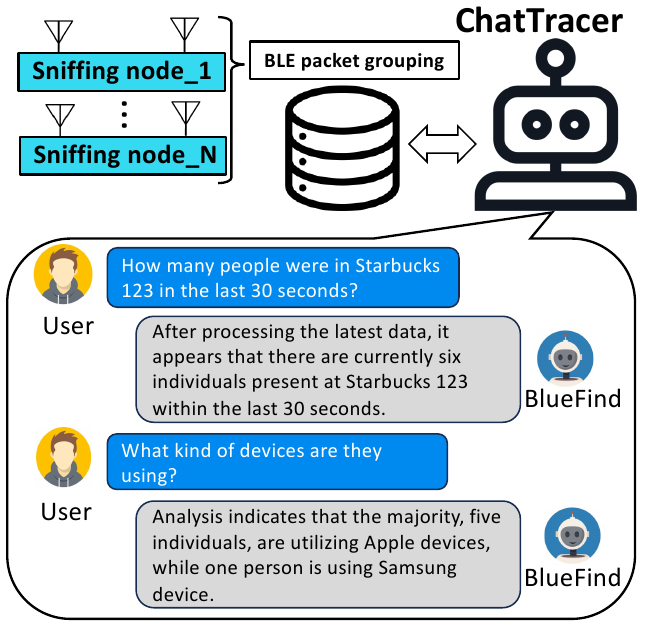} \vspace{-0.1in}
\caption{\pname's system architecture.} \vspace{-0.25in}
\label{fig:systemarc}
\end{figure}

%
Fig.~\ref{fig:systemarc} shows the architecture of \pname. 
It includes a Bluetooth sniffer system, a database, and a fine-tuned LLM.
In what follows, we highlight each component.

\noindent
\textbf{Bluetooth Sniffing Nodes:}
The sniffing system consists of an array of radio receivers distributed in areas of interest such as shopping malls, libraries, and supermarkets. 
Each radio receiver is equipped with multiple antennas. 
The 2.4GHz radio signals are first down-converted to I/Q samples, which are then fed into a polyphase channelizer. 
The channelizer divides the 80MHz channel into 40 2MHz channels, which are processed in parallel.
If a Bluetooth frame is identified on a channel, it will be demodulated to bits sequence.

\noindent
\textbf{Physical-Layer \& Payload Feature Extraction:}
At each sniffing node, the physical-layer features including timestamp, received signal strength (RSS), and carrier frequency offset (CFO) are extracted from each detected BLE Adv packet. Meanwhile, the angle of arrival (AoA)  is estimated for each received packet using the improved MUSIC algorithm \cite{schmidt1986multiple}.

In addition to extracting its physical-layer features, each BLE Adv packet will be decoded to get the message it carries. 
%
If the packet is from an Android device, we can get its advertising address, UUID (equivalently device model), and Google's UUID code. 
If the packet is from an Apple device, we can obtain much more information by decoding its ACMs.

\noindent

\noindent
\textbf{Fine-Tuned LLM:}
\pname employs Mistral-7B \cite{jiang2023mistral} as its LLM. 
Mistral-7B is a foundation model developed by Mistral AI \cite{jiang2023mistral}, supporting a variety of use cases such as text generation, summarization, classification, and text completion.
%
In \pname, Mistral-7B sits behind the database, interacting with end users through human-like text. 


 

 \vspace*{-0.1in}
\section{Implementation}

\textbf{Bluetooth Sniffing Node:}
We implemented the Bluetooth sniffing node for \pname using a BladeRF 2.0 micro xA4 device \cite{bladeRF}, which is a software-defined radio device that supports 61.44MHz sampling rate.
The BladeRF device is equipped with BT-200 Bias-Tee Low Noise Amplifier (LNA) \cite{bt200} and a 12 dBi omnidirectional antenna.
Our measurements indicate that our sniffing node has a 20-meter radius for BLE packet detection.


\textbf{Database and LLM:}
We manage the database using MySQL. 
The LLM (Mistral-7B) was trained (fine-tuned) on a computer server using four Nvidia V100 32GB. 
The trained LLM was then deployed on a MacBook Pro for inference in real time. 
End users can interact with the LLM via the Internet.

\section{Performance Evaluation}

\begin{figure}
     \centering
     \begin{subfigure}[b]{0.15\textwidth}
         \centering
         \includegraphics[width=\textwidth]{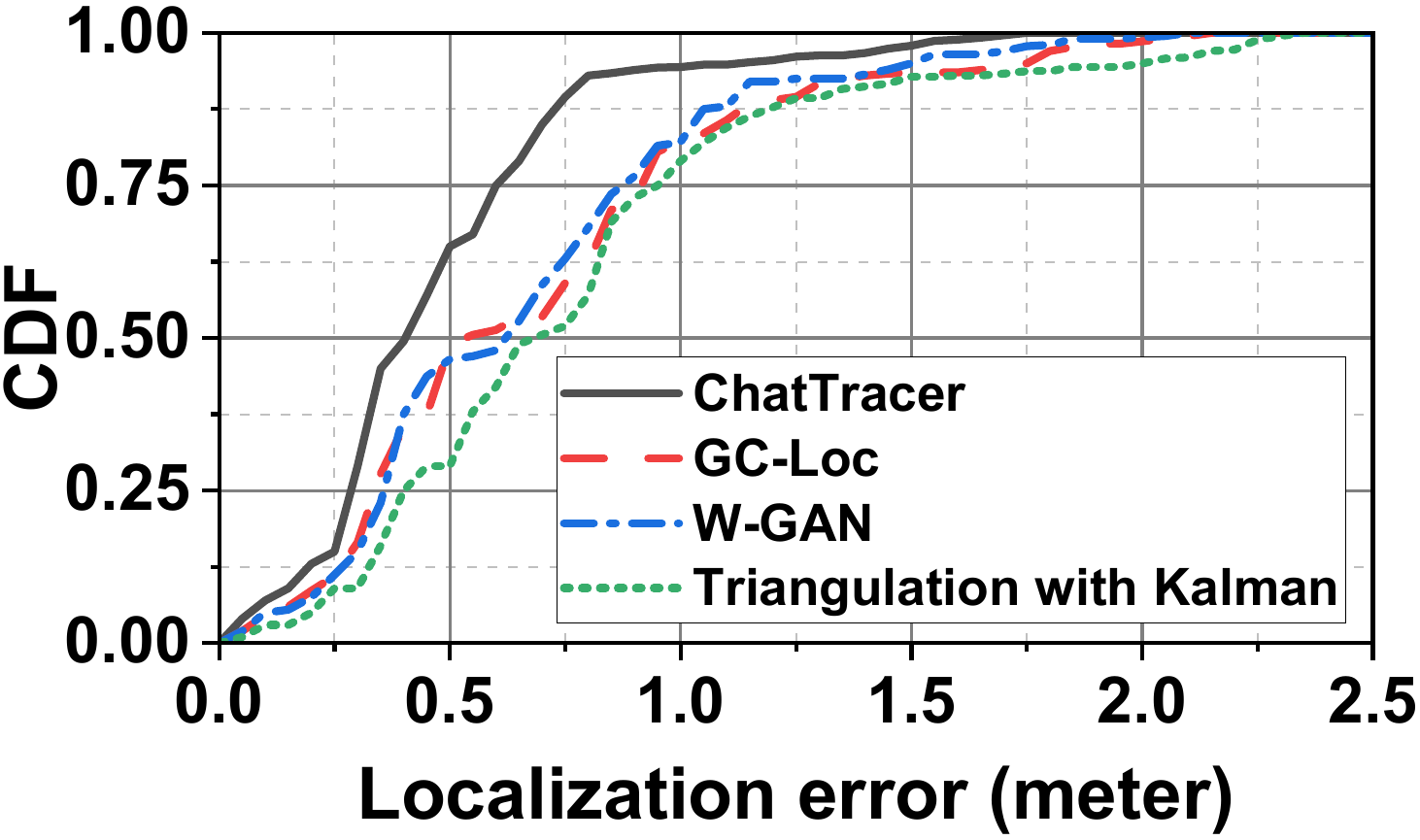}
         \caption{Apartment.}
         \label{fig:apt_cdf}
     \end{subfigure}
     \hfill
     \begin{subfigure}[b]{0.15\textwidth}
         \centering
         \includegraphics[width=\textwidth]{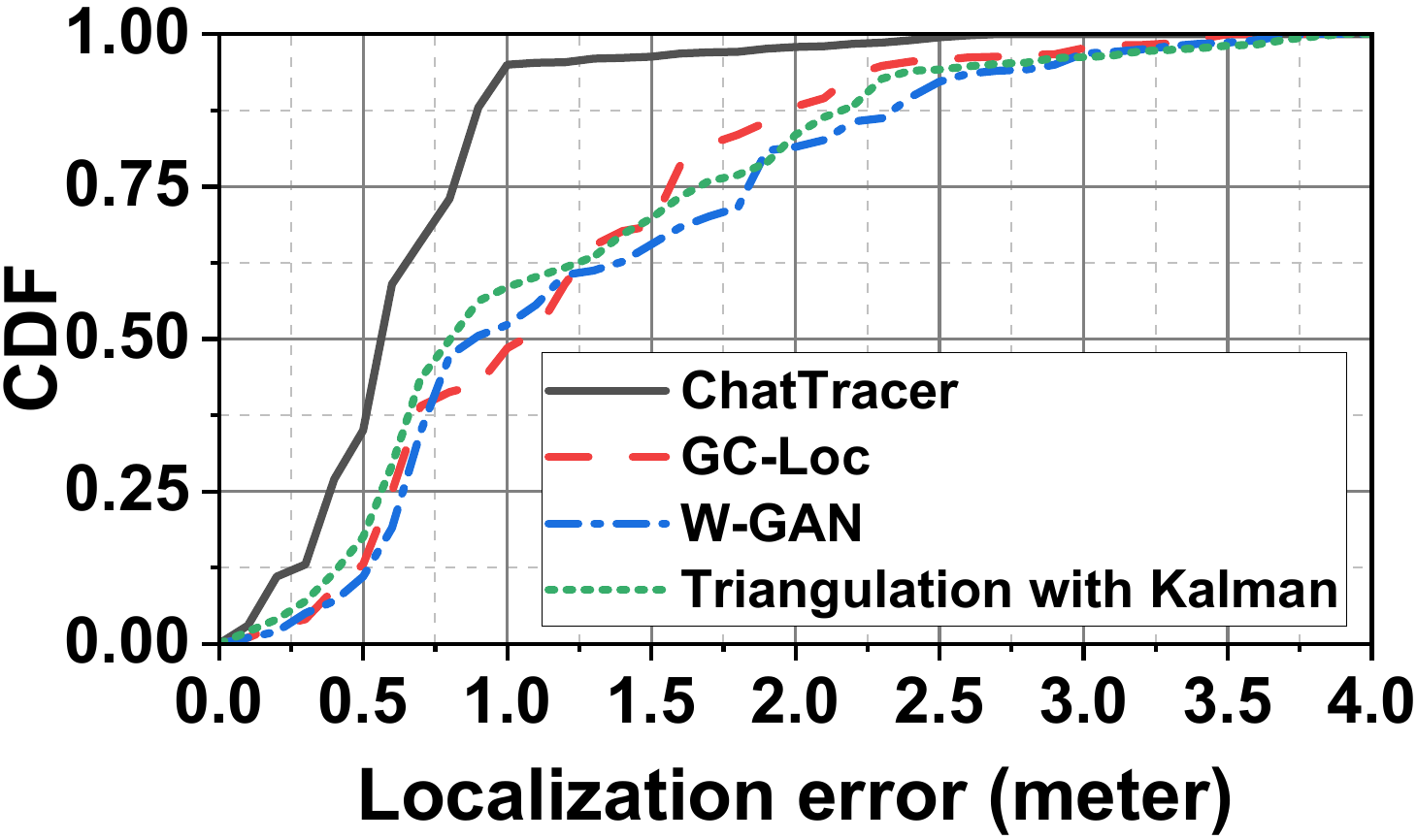}
         \caption{Laboratory.}
         \label{fig:lab_cdf}
     \end{subfigure}
     \hfill
     \begin{subfigure}[b]{0.15\textwidth}
         \centering
         \includegraphics[width=\textwidth]{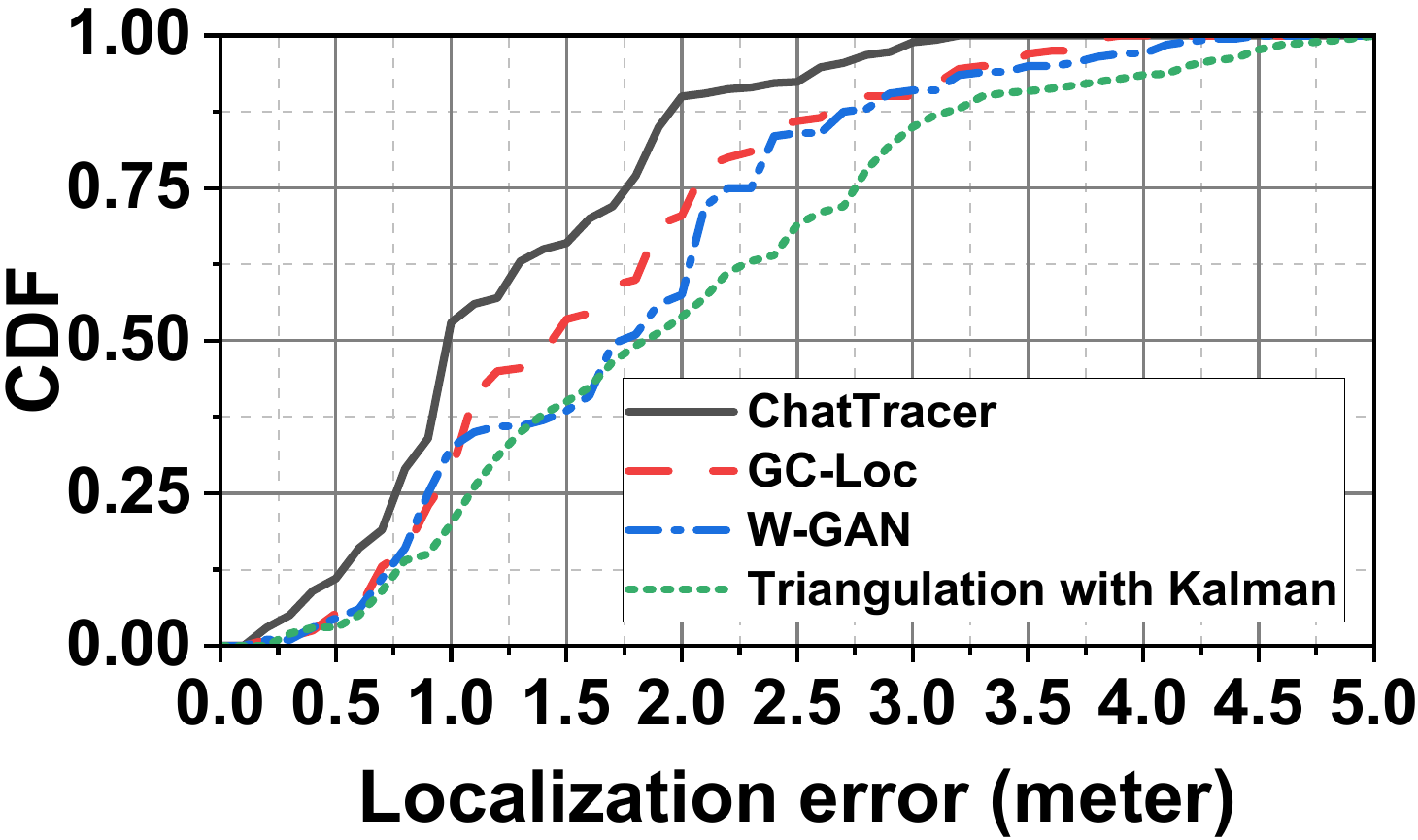 }
         \caption{Shopping mall.}
         \label{fig:mall_cdf}
     \end{subfigure} 
        \caption{Location error distribution of \pname in comparison with model-based localization.}\vspace{-0.2in}
        \label{fig:all_cdf}
\end{figure}

\begin{figure}
     \centering
     \begin{subfigure}[b]{0.15\textwidth}
         \centering
         \includegraphics[width=\textwidth]{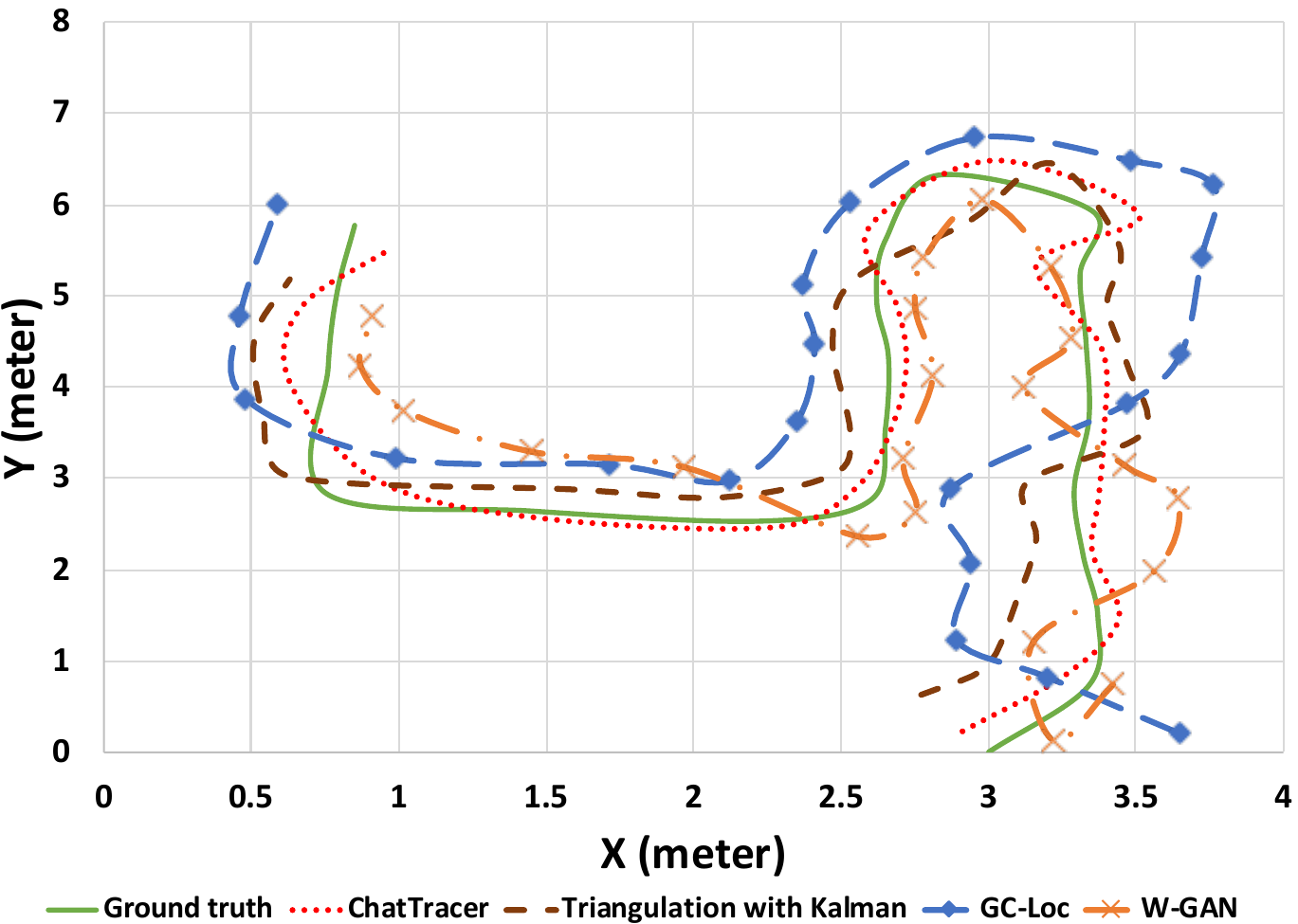}
         \caption{Apartment.}
         \label{fig:apt_traj}
     \end{subfigure}
     \hfill     
     \begin{subfigure}[b]{0.15\textwidth}
         \centering
         \includegraphics[width=\textwidth]{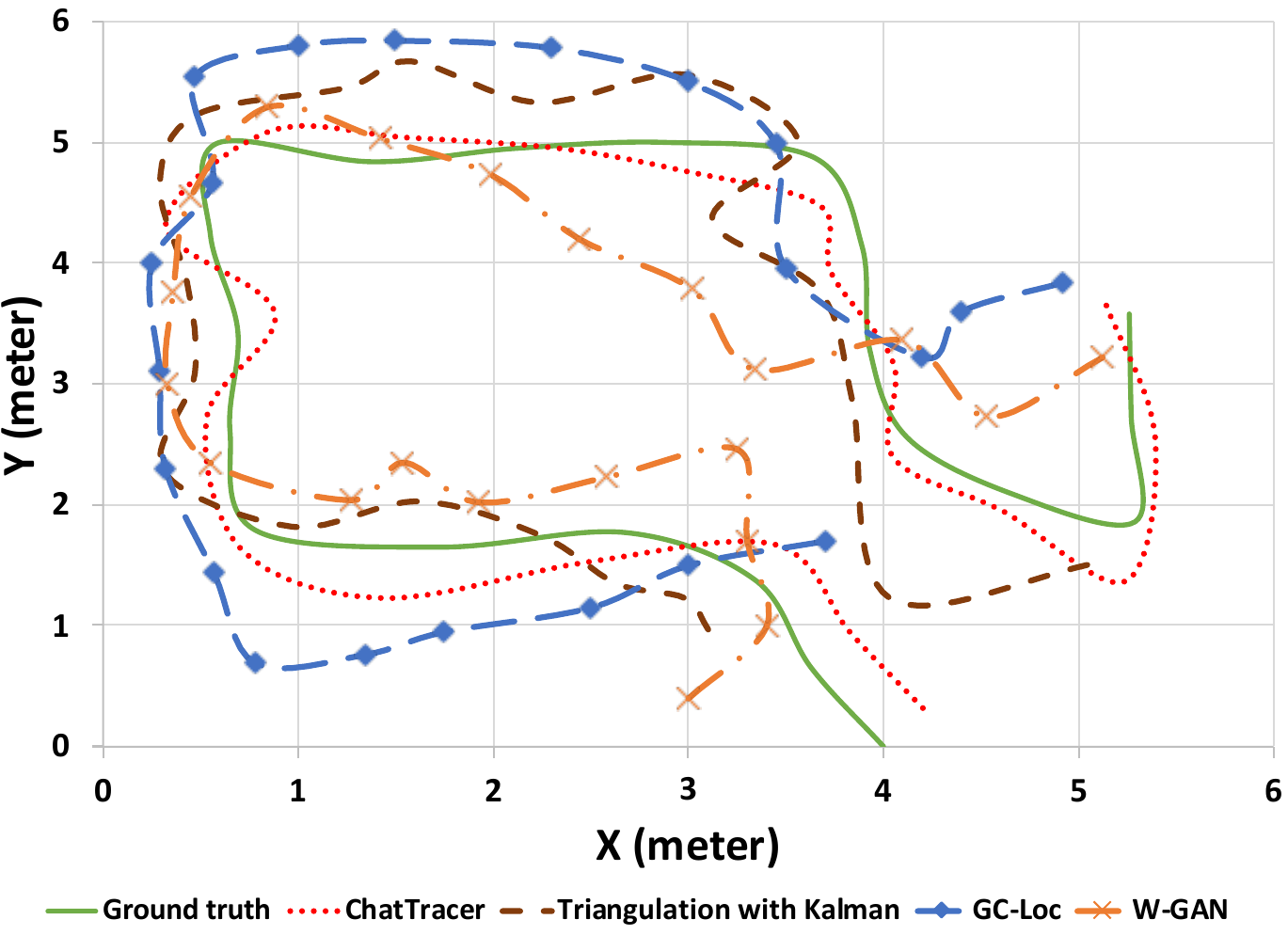}
         \caption{Laboratory.}
         \label{fig:lab_traj}
     \end{subfigure}
     \hfill
     \begin{subfigure}[b]{0.15\textwidth}
         \centering
         \includegraphics[width=\textwidth]{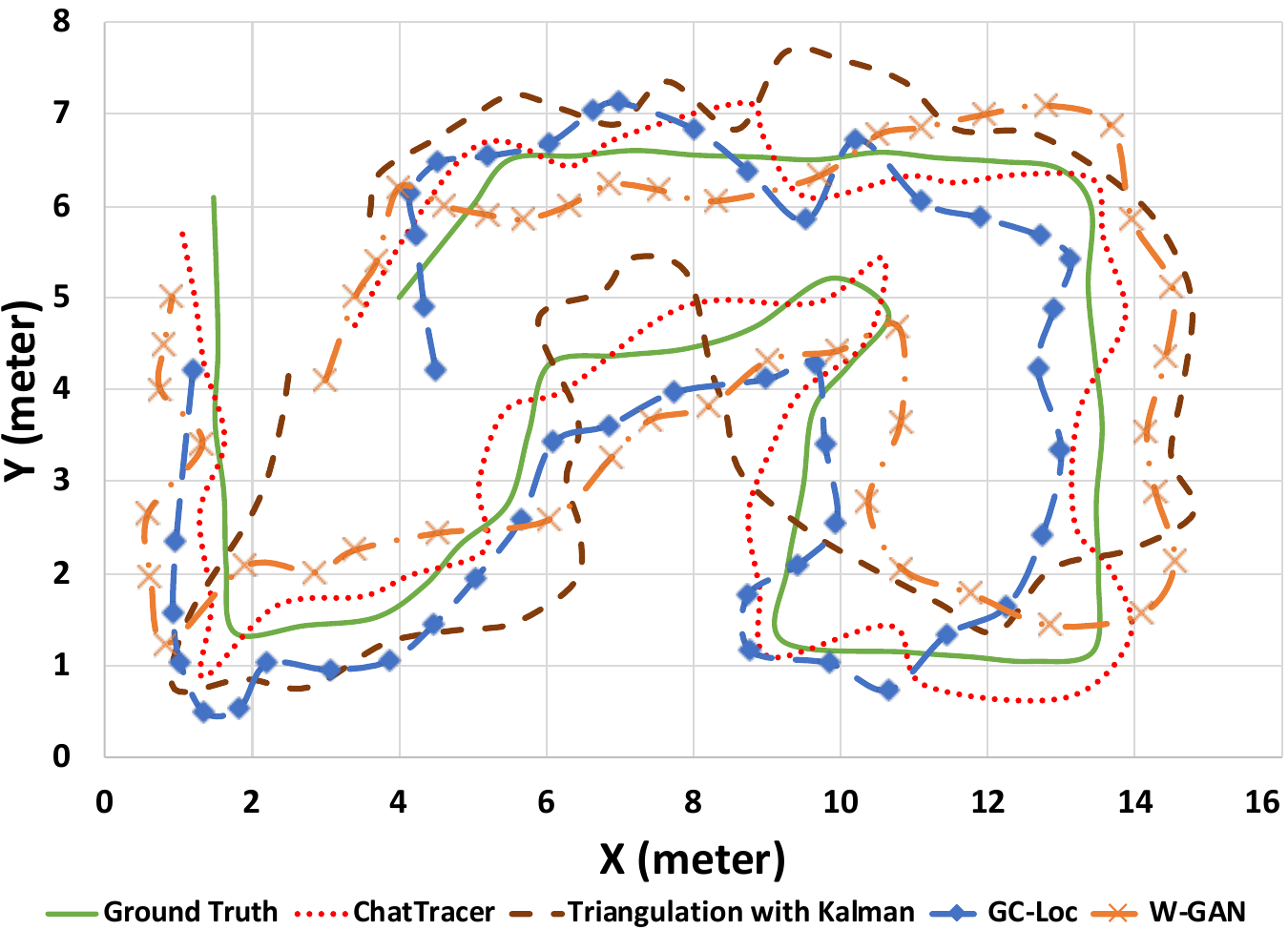}
         \caption{Shopping mall.}
         \label{fig:mall_traj}
     \end{subfigure} 
         \caption{Illustration of \pname's trajectory accuracy in three scenarios.} \vspace{-0.2in}
        \label{fig:all_traj}
\end{figure}

To evaluate the localization accuracy of \pname, we selected 20 locations in the Apartment, 30 locations in the Laboratory, and 50 locations in the Shopping Mall. 
At each location, we measured location errors 300 times for each of the 30 Bluetooth devices. 
Fig.~\ref{fig:all_cdf} presents our measured results. 
It can be seen that \pname outperforms the three SOTA baseline approaches in all three scenarios.
We also conducted experiments to evaluate \pname's ability to track an iPhone within a person's pocket.
Again, we use those three SOTA approaches as our comparison baselines. 
In each of the three scenarios, we marked a path and asked different persons to walk along the path at a speed of 1m/s with iPhones in their pockets.
Fig.~\ref{fig:all_traj} shows the trajectories generated by \pname and its comparison baselines as well as the ground truth.
It is observable that \pname always outperforms its counterparts. 
\section{Conclusion}
This paper studied the possibility of connecting LLM with WSN, aimed at 
i) expanding LLM's knowledge landscape to the physical world,
and
ii) transforming the human interaction with WSN. 
We have built a prototype of \pname and evaluated its performance in three different, realistic scenarios. 
Experimental results confirm the localization accuracy superiority of \pname compared to existing approaches.
Experiments also showcase the new way of information integration and human interaction with WSN.

\bibliographystyle{ACM-Reference-Format}
\bibliography{references}

\end{document}